# The Resonance Condition for Slow Wave Antennas: a Lagrangian Approach

Robert Nevels, Steven Scully, Francisco Espinal, and Anatoly Svidzinsky

*Abstract*— A proof of the resonant property of linear periodically loaded antennas with subwavelength elements is obtained by applying a Lagrangian formalism. A Lagrangian is developed by modeling the antenna with lumped inductance and capacitance elements on a single line, thereby physically similar to the antenna and thus avoiding the inaccurate two parallel conductor transmission line model. An equation for the antenna current driven by an incident electromagnetic field is obtained via vector and scalar potentials. It is shown that periodic loading provides a means to shorten the resonant length while the antenna pattern remains unchanged. The Lagrangian model is validated through a calculation showing the loaded resonant length is determined by a product of resonant dipole length $\lambda/2$ with the ratio of the free space velocity and the longitudinal traveling wave velocity. A periodically loaded disk-on-rod antenna example with simulations and measurements provides further validation of the mathematics.

## I. Introduction

In the present paper we provide a mathematical proof that linear periodically loaded antennas resonate at a length below that of an unloaded resonant antenna, $\lambda/2$, while the radiation pattern remains the same. Some slow wave antennas that take advantage of structural periodicity include engineered composite and bandgap radiators [2], Leaky-wave antennas [3], Yagi and log periodic designs [4], and surface metamaterial constructs [5]. Here we create a discrete model by subdividing a linear antenna into segments characterized in terms of their inductance and capacitance properties. From this information a Lagrangian is formulated and solved using a variational technique [6], [7].

Unlike the standard unloaded antenna, to the authors' knowledge an accurate mathematical derivation of the resonant characteristic of a slow wave antenna has not been previously available. In former work researchers have relied on transmission line models or input impedance characterization at the antenna terminals to demonstrate resonance [8,9,10,11]. While transmission line models can accurately predict the impedance of an antenna, the two parallel wires required for transmission line analysis do not replicate the physical configuration of a dipole, a monopole, or other antenna designs [1-5]. Also, transmission line methods do not provide the correct modes and wavenumbers for a loaded or even an unloaded linear antenna. Here we introduce a Lagrangian method which we show is uniquely suited for modeling loaded linear antenna designs. An outcome of this study is a proof that periodically loaded antennas have a resonant condition that is a simple modification of the unloaded case and a field pattern, which is the same as that of the loaded case at resonance yet occurs at a lower resonant frequency. Below we first review the classic formulation for the resonant condition of an unloaded dipole and then develop the mathematical model and solution for loaded dipole antennas.

## II. Formulation

Mathematical investigations of the resonant characteristic of the unloaded linear dipole are found in seminal work by work by Hallén [12], Schelkunoff [13], and King [14]. However, in [15] Leontovich and Levin offered the definitive mathematical derivation of the resonance condition for a linear unloaded dipole antenna. For completeness, we begin with a brief description of the Leontovich and Levin derivation.

### A. Dipole resonance analytical method

The Leontovich and Levin method requires applying thin linear conductor conditions on the vector potential integral equation and hypothesizing a method of perturbations power series approximation for the antenna current

$$I(z) = I_0(z) + I_1(z)\chi + I_2(z)\chi^2 + \cdots \quad (1)$$

where the subscripted $I's$ are coefficients of $\chi = 1/(2\ln(ka))$, $a$ is the antenna radius, and $k$ the free space wavenumber. This expression for the current, when substituted into an approximated vector potential equation, results in a differential equation for each $I_n$ coefficient. For example, when the lowest order coefficient equation,

This work was supported by the Air Force Office of Scientific Research (Award FA9550-20-1-0366 DEF), the Office of Naval Research (Awards N00014-20-1-2184 and N00014-16-1-2578), the Robert A. Welch Foundation (Award A-1261), the National Science Foundation (Award No. PHY-2013771), and the Natural Science Foundation of Fujian (Award No. 2021I0025).

R. Nevels and F. Espinal are with the Electrical and Computer Engineering Department, Texas A&M University, College Station, TX, 77843 USA (e-mail: r-nevels@tamu.edu).

S. Scully is with Wavetrix Inc., 300 Municipal Dr., Richardson, TX, 75080 (e-mail: sscully@wavetrix.com).

A. Svidzinsky is with the Department of Physics & Astronomy, Texas A&M University, College Station, TX, 77843 USA and School of Optoelectronic and Communication Engineering, Xiamen University of Technology, Xiamen 361024, China (e-mail: asvid@physics.tamu.edu).

$$\frac{\partial^2 I_0}{\partial z^2} + k^2 I_0 = 0 \tag{2}$$

is subjected to the boundary conditions $I_0(\pm W) = 0$ at the extremities $(z = \pm W)$ of the dipole, no matter how the incident field is distributed over the dipole, the current distribution is,

$$I_0 = i_0 \begin{cases} \cos(kz), n\ odd \\ \sin(kz), n\ even \end{cases}. \tag{3}$$

The fundamental mode angular frequency of oscillation of the current on an unloaded linear antenna is therefore [16]

$$\omega_a = \pi v/L, \tag{4}$$

where $v$ is the speed of propagation of the current wave along the antenna and $L = 2W$ is its length. The angular frequency $\omega = 2\pi c/\lambda$ substituted into (4) at resonance gives $L = \frac{\lambda}{2}\frac{v}{c}$ where $c$ is the velocity of the incident field in air. On the metal dipole both the velocity of the current $v$ and the wavelength $\lambda$ are that of air. Therefore, the dipole length is,

$$L = \lambda/2. \tag{5}$$

We now proceed to derive the resonance condition of a slow wave loaded antenna.

## B. Electric current propagation on a slow wave antenna

We begin by deriving an equation for the current $I(t,z)$ on a loaded antenna by means of the antenna model shown in Fig. 1. The model is a segmented chain of subwavelength physical reactive elements. A cell in the model is composed of a capacitance and the section of the antenna between two consecutive capacitive elements represented by an inductance. The capacitance due to the charge denoted by $Q_i(t)$ that has passed through the $i^{\text{th}}$ inductor of the chain after time $t$, enables a Lagrangian description of the antenna,

$$L_a = \frac{1}{2}\sum_i L\left(\frac{\partial Q_i(t)}{\partial t}\right)^2 - \frac{1}{2C}\sum_i [Q_{i+1}(t) - Q_i(t)]^2, \tag{6}$$

where $L$ and $C$ are the lumped inductance and capacitance, respectively, of the elements that make up the $i^{\text{th}}$ section of the chain. The charge stored in the $i^{\text{th}}$ capacitor is $Q_{i+1}(t) - Q_i(t)$.

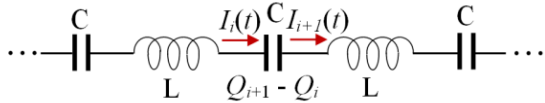

Fig. 1. Model of an antenna with periodic slow wave structures.

We now generalize from a discrete to a continuous system. This is accomplished by rewriting (6) in the form

$$L_a = \frac{1}{2}\int dz\left[l\left(\frac{\partial Q(t,z)}{\partial t}\right)^2 - \frac{1}{c}\left(\frac{\partial Q(t,z)}{\partial z}\right)^2\right], \tag{7}$$

where lower case $l$ and $c$ are the inductance and the capacitance per unit length.

By coupling the antenna current and the incident electromagnetic field, we obtain the following total action of the system,

$$S = \int dtdz\left[\frac{l}{2}\left(\frac{\partial Q(t,z)}{\partial t}\right)^2 - \frac{1}{2c}\left(\frac{\partial Q(t,z)}{\partial z}\right)^2 \right. \\ \left. + I(t,z)A_z(t,z) + \frac{\partial Q(t,z)}{\partial z}\phi(t,z)\right], \tag{8}$$

where $I(t,z) = \partial Q(t,z)/\partial t$ is the electric current on the antenna which lies along the z-axis, $A_z(t,z)$ is the z-component of the incident field vector potential and $\phi(t,z)$ is the incident field scalar potential. The action (8) is invariant under gauge transformation of the scalar and vector potentials $\phi \to \phi + \frac{\partial f}{\partial t}$, $A_z \to A_z - \frac{\partial f}{\partial z}$, where $f(t,z)$ is an arbitrary function.

The equations of motion are obtained in accordance with the principle of least action by varying $S$ with respect to $Q$. We have

$$\delta S = \int dtdz\left[l\frac{\partial Q(t,z)}{\partial t}\frac{\partial \delta Q(t,z)}{\partial t} - \frac{1}{c}\frac{\partial Q(t,z)}{\partial z}\frac{\partial \delta Q(t,z)}{\partial z} \right. \\ \left. + \frac{\partial \delta Q(t,z)}{\partial t}A_z(t,z) + \frac{\partial \delta Q(t,z)}{\partial z}\phi(t,z)\right].$$

By integrating by parts, and taking into account that in accord with the principle of least action at the limits of the time integration, the variation of the charge $\delta Q(t,z)$ is equal to zero, we obtain

$$\delta S = -\int dtdz\left[l\frac{\partial^2 Q(t,z)}{\partial t^2} - \frac{1}{c}\frac{\partial^2 Q(t,z)}{\partial z^2} + \frac{\partial A_z(t,z)}{\partial t} \right. \\ \left. + \frac{\partial \phi(t,z)}{\partial z}\right]\delta Q(t,z).$$

Since according to the principle of least action, the variations $\delta Q$ are arbitrary, the coefficient of the $\delta Q$ must be set equal to zero which yields the equation of motion for the charge on the antenna

$$\frac{\partial^2 Q}{\partial t^2} - v^2\frac{\partial^2 Q}{\partial z^2} = -\frac{1}{l}\frac{\partial A_z}{\partial t} - \frac{1}{l}\frac{\partial \phi}{\partial z} = \frac{1}{l}E_z, \tag{9}$$

where $v = 1/\sqrt{lc}$ is the phase velocity of the charge wave and $E_z$ is the z-component of the electric field $E_z = -\frac{\partial A_z}{\partial t} - \frac{\partial \phi}{\partial z}$. The equation of motion for the electric current on the antenna is found by differentiating (9) over time, yielding

$$\frac{\partial^2 I}{\partial t^2} - v^2\frac{\partial^2 I}{\partial z^2} = \frac{1}{l}\frac{\partial E_z}{\partial t}. \tag{10}$$

In Appendix A, (10) is derived without the driving term by using Kirchhoff's laws to obtain the validity conditions for the continuous system approximation, namely, spacing between antenna elements should be much smaller than the wavelength of the current wave.

*C. Loaded slow wave dipole resonance analytical method*

We now move forward with a one-dimensional wave equation with a driving term, (10), subject to the boundary conditions

$$I(t, -W) = I(t, W) = 0.$$

Equations (10) and (11) model an antenna of length $2W$ driven by electromagnetic wave with an incident electric field $\mathbf{E}$ polarized along the z-axis. We hypothesize that

$$E_z = E_0 \cos(\omega t) \tag{12}$$

and look for solutions of (10) in the form $I(t, z) = \sin(\omega t) I(z)$. Given the time harmonic nature of the incident field, which is mirrored by the current, (10) can easily be converted to the following frequency domain equation for $I(z)$

$$\frac{\partial^2 I(z)}{\partial z^2} + k^2 I(z) = P, \tag{13}$$

where $k = \omega/v$ and $P = \omega E_0/(v^2 l)$, are subject to the boundary conditions

$$I(-W) = I(W) = 0. \tag{14}$$

One can solve (13) by writing a general solution with the form

$$I(z) = \frac{P}{k^2} + D_1 \cos(kz) + D_2 \sin(kz). \tag{15}$$

The arbitrary constants $D_1$ and $D_2$ are obtained by applying the boundary conditions in (14) on (15), which yields

$$\frac{P}{k^2} + D_1 \cos(kW) - D_2 \sin(kW) = 0, \tag{16}$$

$$\frac{P}{k^2} + D_1 \cos(kW) + D_2 \sin(kW) = 0, \tag{17}$$

Simultaneous solution of these equations produces $D_2 = 0$ and $D_1 = -P/(k^2 \cos(kW))$. Therefore,

$$I(z) = \frac{P}{k^2}\left[1 - \frac{\cos(kz)}{\cos(kW)}\right]. \tag{18}$$

Equation (18) shows that current amplitude becomes large when $\cos(kW) = 0$. This yields the following formula for the resonant length of the antenna

$$L = \frac{2\pi\left(n + \frac{1}{2}\right)}{k} = \lambda_0\left(n + \frac{1}{2}\right)\frac{v}{c} \tag{19}$$

where $n = 0, 1, 2, \ldots$ and $L=2W$, the length of the antenna, $v$ is the traveling wave velocity, and $c$ is the free space wave velocity. For the fundamental mode, $n = 0$, (19) reduces to

$$L = \frac{\lambda}{2}\frac{v}{c}. \tag{20}$$

Equation (20) shows that the resonant length of the loaded resonant antenna is proportional to $v/c$. Therefore, to shorten a loaded resonant antenna one should reduce the speed of the current. Since $v = 1/\sqrt{lc}$ this can be achieved by changing the capacitance or inductance of the loaded antenna elements [17,18]. As an example of a periodic traveling wave resonant antenna, below we present a disk-on-rod design but with elements that are significantly subwavelength in size.

## III. SIMULATION AND EXPERIMENT

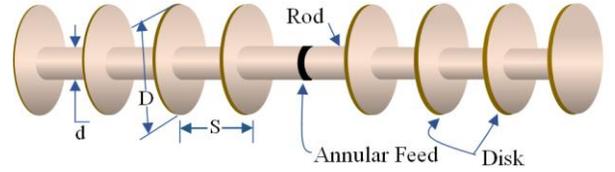

Fig. 2 Rod and disk configuration with rod diameter d, disk diameter D, and disk separation distance S. The rod supplies the inductance and the disks the capacitance.

To test the hypothesis above, a disk and rod configuration illustrated in Fig. 2 was chosen because of its rotational symmetry and the ease with which the antenna capacitance can be adjusted by changing the radius of the disks. A numerical calculation of reflection coefficient versus frequency for disk radii from D/2 = 3 mm to 21 mm is shown in Fig. 3 in terms of the scattering matrix coefficient $S_{11}$. The rod contains 8 disks and maintains a length is 137 mm and diameter of 4 mm. The capacitance due to facing disks is given by $C = \varepsilon A/S$, where $\varepsilon$, $A$, and $S$ are respectively the permittivity of air, the facing area of the disks, and the separation distance between the disks. The simulation in Fig. 3 shows that as the radius of the disks increases thereby increasing the facing area A, the disks capacitance increases and therefore the resonant frequency of the of the antenna decreases. An optimum impedance match condition occurs at approximately 850 MHz where the disk radius is D/2 = 6 to 7 mm. This means that the disks are significantly sub-wavelength, extending only 2 to 3 mm above the rod surface, but none-the-less this is where the inductances and capacitances balance to form an ideal resonance. The distance S between the plates and the rod diameter do not change and therefore the inductance remains constant leaving a narrow band of frequencies where the inductance and capacitance create ideal resonance. The resonance of the unloaded dipole, the right most curve in Fig. 3 occurs at 1 GHz and has an $S_{11}$ of -15 dB. As the disk radii increase, resonance shifts down in frequency and the bandwidth becomes narrower, 325 MHz for the unloaded dipole as compared to 225 MHz for similar $S_{11}$ at 675 MHz. Numerical simulations were made with COMSOL© and HFSS© commercial software.

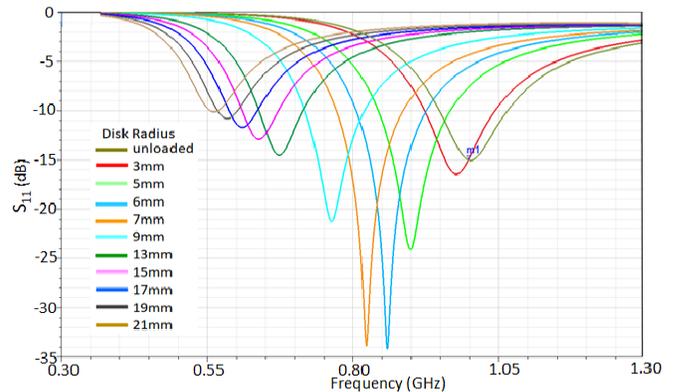

Fig. 3. Simulated $S_{11}$ for nine disk radii. Optimum L-C resonant response occurs where the disk radius is between 6 and 7 mm with $S_{11}$ around -34 dB.

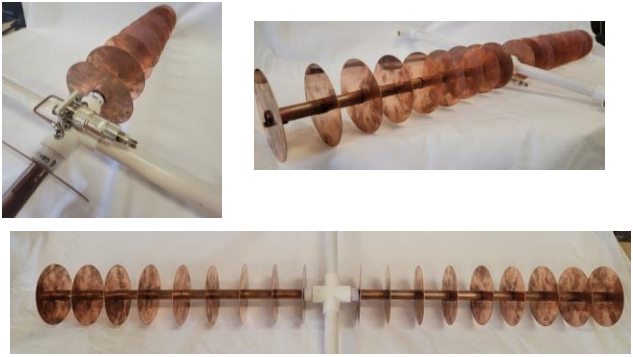

Fig. 4. Views of the loaded measured dipole antenna with sub-wavelength capacitive plate disks and rod inductance used to perform antenna pattern and impedance measurements.

In order to maintain consistent input feed and rod construction a commercially available dipole was purchased for our measurements, shown with disk loading added in Fig. 4. The copper tube rod diameter is d = 19 mm and the disks are milled from 24-gauge copper sheet. The dipole length is 86.5 cm. Measurements were performed in ETS Lindgren anechoic chamber which can make high fidelity measurements from 800 MHz to 120 GHz with a Keysight Technologies network analyzer.

Figure 5 shows the E-plane pattern is on the left and the H-plane pattern on the right for the bare dipole and 38mm and 48 mm diameter disk loading. Each measurement is taken at the first resonance of the respective antennas. As can be seen there is very little overall difference between the three antenna radiation patterns.

## IV. CONCLUSIONS

The resonant condition for a dipole antenna periodically loaded with subwavelength elements has been derived through a combination of Lagrangian and classical mathematical methods. The model, which relies on an incident electromagnetic field expressed in terms of the vector and scalar potentials, provides the Lagrangian [6]. The Lagrangian model is linear and therefore physically like the linear antenna it models, as opposed to the more commonly used two wire transmission line model. We obtain the classical action for the a discrete distribution of antenna elements and in the continuous limit, which is valid because the size and spacing between elements are much smaller than the wavelength. Variation of the action leads to the equation of motion for the electric current. The final expression for the eigenvalues for the resonant modes then follows by applying the boundary conditions on the current.

The main contribution of this paper is the derivation of (18), which yields the resonance condition (19). To the authors' knowledge there is no previous derivation of (18). Also, through (18), we have shown that the length of a resonant traveling wave antenna is proportional to the ratio of the phase velocity of the antenna current to the free space velocity, and therefore the resonant antenna length of a slow wave antenna can be much less than the resonant length of an unloaded antenna.

As an example confirming our analysis, simulations and measurements were performed on a rod and disk antenna. This antenna structure was chosen for its clearly identifiable rotationally symmetric inductive and capacitive components and the flexibility of changing the capacitance by adjusting the disk diameter, thereby allowing resonance condition tuning capability. The rod and disk simulated $S_{11}$ confirms the resonant condition (20) and the measured radiation patterns in Fig 5 show the radiation pattern at resonance is virtually independent of the resonant frequency brought about by changes in the antenna loading.

## APPENDIX

*Antenna model with metal sphere structures*

Here we derive the propagation equation (10) for the electric current in a slow wave antenna by assuming the capacitive structures in the antenna model, Fig. 1, are metal spheres of radii $R$. If the sphere radius is small compared to the spacing between the spheres the potential of sphere $i$ is mainly determined by the sphere charge $q_i$

$$V_i = \frac{q_i}{4\pi\varepsilon_0 R} = \frac{Q_{i+1}(t) - Q_i(t)}{4\pi\varepsilon_0 R}. \quad (A1)$$

The voltage between spheres $i+1$ and $i$ can be written as

$$V_{i+1} - V_i = L\frac{\partial I_{i+1}}{\partial t}. \quad (A2)$$

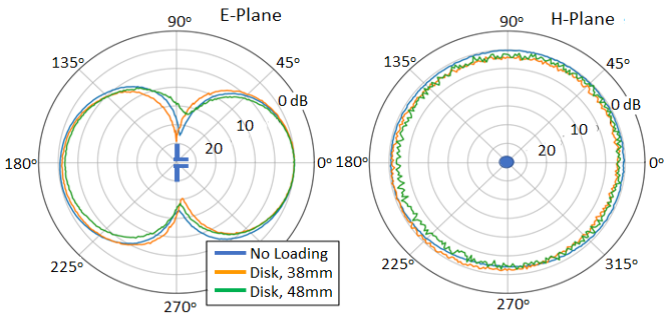

Fig. 5. Measured dipole and disk loaded dipole at first resonance E-plane and H-plane patterns.

Combining (A1) and (A2), and taking into account $I_{i+1} = \frac{\partial Q_{i+1}}{\partial t}$, we obtain

$$\frac{\partial^2 I_i}{\partial t^2} - \frac{I_{i+1} - 2I_i + I_{i-1}}{LC} = 0, \quad (A3)$$

where $C = 4\pi\varepsilon_0 R$ is a self-capacitance of a conducting sphere of radius $R$. The second order spatial derivative can be approximated as

$$\frac{\partial^2 I_i}{\partial z^2} \approx \frac{I_{i+1} - 2I_i + I_{i-1}}{d^2}, \quad (A4)$$

where $d$ is the spacing between spheres. With (A4), (A3) can be written as a propagation equation

$$\frac{\partial^2 I}{\partial t^2} - v^2 \frac{\partial^2 I}{\partial z^2} = 0, \quad (A5)$$

where $v = d/\sqrt{LC} = 1/\sqrt{lc}$, $l = L/d$ and $c = C/d$ are the inductance and the capacitance per unit length respectively.

To find the condition of validity of the continuous approximation (A4), we compare the solution of the discrete equation (A3) with a traveling wave solution of the continuous equation (A5) which has the form $I = I_0 e^{-j\omega t + jkz}$, where

$$\omega = vk. \quad (A6)$$

We look for a solution of the discrete equation (A3) in the form $I_i = I_0 e^{-j\omega t + j\xi i}$. By inserting this current expression into (A3), we obtain the following dispersion relation

$$\omega^2 = \frac{2}{LC}[1 - \cos(\xi)]. \quad (A7)$$

The dispersion relation (A7) reduces to (A6) in the limit $\xi \ll 1$. In this limit

$$\omega \approx \frac{v\xi}{d} \approx vk, \quad (A8)$$

where $v = d/\sqrt{LC}$. Thus, the continuous approximation is valid provided $d \ll v/\omega$, that is spacing between antenna elements should be much smaller than the wavelength of the current wave.